\documentclass[12pt,epsf,epsfig,psfig]{article}
\usepackage{mathptmx}        
\usepackage{amssymb,amsmath}
\usepackage{graphicx}
\usepackage{epsfig}
\usepackage{cite}
\newcommand{\nc}{\newcommand}
\nc{\req}[1]{Eq.\,(\ref{#1})}    \nc{\reqp}[1]{Eq.\,(\ref{#1}) on page \pageref{
#1}}
\nc{\rf}[1]{Fig.~\ref{#1}}   \nc{\rfp}[1]{Fig.~\ref{#1} on page \pageref{#1}}

\nc{\Th}{\ensuremath{T_\mathrm{H}\,}}
\nc{\pp}{\ensuremath{p\!p\ }}
\nc{\pA}{\ensuremath{p A\ }}
\nc{\hAA}{\ensuremath{AA\ }}
\def\beq{\begin{equation}}
\def\eeq{\end{equation}}

\def\e{\epsilon}

\def\pp{\cal pp}
\def\pA{\cal pA}
\def\hAA{\cal AA}

\def\be{\begin{equation}}
\def\ee{\end{equation}}

\def\lsim{\raise0.3ex\hbox{$<$\kern-0.75em\raise-1.1ex\hbox{$\sim$}}}
\def\gsim{\raise0.3ex\hbox{$>$\kern-0.75em\raise-1.1ex\hbox{$\sim$}}}

\begin{document}

January 2015 \hfill BI-TP 2015/09

\vskip 3cm

\centerline{\Large \bf The Legacy of Rolf Hagedorn:}

\vskip0.5cm

\centerline{\Large \bf  Statistical Bootstrap and Ultimate Temperature$^*$}

\vskip 1.5cm

\centerline{\large  Krzysztof Redlich$^1$
and Helmut Satz$^2$ }

\vskip 0.5cm

\medskip

\centerline{$^1$ Institute of Theoretical Physics, University of Wroc{\l}aw,
Poland}

\medskip

\centerline{$^2$ Fakult\"at fur Physik, Universit\"at Bielefeld, Germany}

\vskip 2cm

\centerline{\large \bf Abstract:}

\bigskip

In the latter half of the last century, it became evident that there
exists an ever increasing number of different states of the so-called
elementary particles. The usual reductionist approach to this problem
was to search for a simpler infrastructure, culminating in the
formulation of the quark model and quantum chromodynamics. In a complementary,
completely novel approach, Hagedorn suggested that the mass distribution of
the produced particles follows a self-similar composition pattern, predicting
an unbounded number of states of increasing mass. He then concluded that such a
growth would lead to a limiting temperature for strongly interacting matter.
We discuss the conceptual basis for this approach, its relation to critical
behavior, and its subsequent applications in different areas of high energy
physics.

\vfill

\hrule

\medskip

$^*$ To appear in {\sl Melting Hadrons, Boiling Quarks}, R.\ Hagedorn and
J.\ Rafelski (Ed.), Springer Verlag 2015, open access.

\newpage

\hfill {\sl A prophet is not without honour,}

\hfill {\sl but in his own country.}

\medskip

\hfill The New Testament, Mark 6,4.

\vskip1cm

{\large{\section{Rolf Hagedorn}}}

The development of physics is the achievement of physicists, of humans,
persisting against often considerable odds. Even in physics, fashion rather
than fact is frequently what determines the judgement and recognition.

\medskip

When Rolf Hagedorn carried out his main work, now quite generally recognized
as truly pioneering, much of the theoretical community not only ignored it,
but even
considered it to be nonsense. ``Hagedorn ist ein Narr'', he is a fool, was
a summary of many leading German theorists of his time. When in the 1990's
the question was brought up whether he could be proposed for the Max Planck
Medal, the highest honor of the German physics community, even then, when
his achievements were already known world-wide, the answer was still
``proposed, yes...'' At the time Hagedorn carried out his seminal research,
much of theoretical physics was ideologically fixed on ``causality, unitarity,
Poincar\'e invariance'': from these three concepts, from axiomatic quantum
field theory, all that is relevant to physics must arise. Those who thought
that science should progress instead by comparison to
experiment were derogated as ``fitters and plotters''. Galileo was almost
forgotten...  Nevertheless, one of the great Austrian theorists of the time,
Walter Thirring, himself probably closer to the fundamentalists, noted:
``If you want to do something really {\sl new}, you first have to have a
{\sl new idea}''.  Hagedorn did.

\medskip

He had a number of odds to overcome. He had studied physics in G\"ottingen
under Richard Becker, where he developed a life-long love for thermodynamics.
When he took a position at CERN, shortly after completing his doctorate, it
was to perform calulations for the planning and construction of the proton
synchrotron. When that was finished, he shifted to the study of multihadron
production in proton-proton collisions and to modelling the results of these
reactions. It took a while before various members of the community, including
some of the CERN Theory Division, were willing to accept the significance of
his work. This was not made easier
by Hagedorn's strongly focussed region of interest, but eventually it became
generally recognized that here was someone who, in this perhaps a little
similar to John Bell, was developing truly novel ideas which at first sight
seemed quite specific, but which eventually turned out to have a lasting
impact also on physics well outside its regions of origin.

\medskip

We find that Rolf Hagedorn's work centers on two themes:

\begin{itemize}
\vskip0.2cm
\item{the statistical bootstrap model, a self-similar scheme for the
composition and decay of hadrons and their resonances; for Hagedorn,
these were the ``fireballs''.}
\item{the application of the resulting resonance spectrum in an ideal gas
containing all possible hadrons and hadron resonances, and to the
construction of
hadron production models based on such a thermal input.}
\end{itemize}
We will address these topics in the first two sections, and then turn to
their role both in the thermodynamics of strongly interacting matter and in
the description of hadron production in elementary as well as nuclear
collisions. Our aim here is to provide a general overview of Hagedorn's
scientific achievements; other aspects will be covered in other chapters of
this book. Some of what we will say transcends Hagedorn's life. But then,
to paraphrase Shakespeare, we have come to praise Hagedorn, not to bury
him; we want to show that his ideas are still important and very much alive.

{\large{\section{The Statistical Bootstrap}}}

Around 1950, the world still seemed in order for those looking for the
ultimate constituents of matter in the universe. Dalton's atoms had been
found to be not really {\sl atomos}, indivisible; Rutherford's model of
the atom had made them little planetary systems, with the nucleus as the
sun and the electrons as encircling planets. The nuclei in turn consisted
of positively charged protons and neutral neutrons as the essential mass
carriers. With an equal number of protons and electrons, the resulting
atoms were electrically neutral, and the states obtained by considering
the different possible nucleus compositions reproduced the periodic table
of elements. So for a short time, the Greek dream of obtaining the entire
complex world by combining three simple {\sl elementary} particles
in different ways seemed finally feasible: protons,
neutrons and electrons were the building blocks of our universe.

\medskip

But there were those who rediscovered an old problem, first formulated
by the Roman philosopher Lucretius: if your elementary particles, in our
case the protons and neutrons, have a size and a mass, as both evidently
did, it was natural to ask what they are made of. An obvious way to
find out is to hit them against each other and look at the pieces. And it
turned out that there were lots of fragments, the more the harder the
collision.
But they were not really pieces, since the debris found after a proton-proton
collision still also contained the two initial protons. Moreover, the
additional fragments, mesons and baryons, were in almost all ways as
elementary as protons and neutrons.
The study of such collisions was taken up by more and more laboratories
and at ever higher collision energies. As a consequence, the number of
different ``elementary'' particles grew by leaps and bounds, from tens to
twenties to hundreds. The latest compilation of the Particle Data Group
contains over a thousand.

\medskip

What to do? One approach was in principle obvious: just as Dalton's atoms
could be
constructed from simpler, more elementary constituents, so one had to find
a way of reproducing all the hadrons, the particles formed in strong
interaction collisions, in terms of fewer and more elementary building
blocks. This conceptually straight-forward problem was, however, far from
easy: a simply additive composition was not possible. Nevertheless, in the
late 1960's the quark model appeared, in which three quarks and their
antiquarks were found to produce in a non-Abelian composition all the
observed states and more, predicted and found. Not much later,
quantum chromodynamics (QCD) appeared as the quantum field theory governing
strong interactions; moreover, it kept the quarks inherently confined,
without individual existence: they occurred only as quark-antiquark pair
(a meson) or as a three-quark state (a baryon). And wealth of subsequent
experiments confirmed QCD as the basic theory of strong interactions.
The conventional reductionist approach had triumphed once more.

\medskip

Let us, however, return to the time when physics was confronted by all those
elementary particles, challenging its practioners to find a way out. At this
point, in the mid 1960's, Rolf Hagedorn came up with a truly novel idea
\cite{rh2} - \cite{Hagedorn3}.
He was not so much worried about the specific properties of the particles.
He just imagined that a heavy particle was somehow composed out of lighter
ones, and these again in turn of still lighter ones, and so on, until one
reached the pion as the lightest hadron. And by combining heavy ones, you
would get still heavier ones, again: and so on. The crucial input was that
the composition law should be the same at each stage. Today we call that
self-similarity, and it had been around in various forms for many years.
A particularly elegant formulation was written a hundred years before
Hagedorn by the English mathematician Augustus de Morgan, the first president
of the London Mathematical Society:

\bigskip

{\sl
\centerline{Great fleas have little fleas upon their backs to bite'em,}

\centerline{and little fleas have lesser still, and so ad infinitum.}

\centerline{And the great fleas themselves, in turn, have greater fleas to
go on,}

\centerline{while these again have greater still, and greater still,
and so on.}
}
\bigskip

Hagedorn proposed that ``a fireball consists of fireballs, which in turn
consist of fireballs, and so on...''  The concept later reappeared in various
forms in geometry; in 1915, it led to the celebrated triangle devised by
the Polish mathematician Wac{\l}aw Sierpinski: `` a  triangle
consists of triangles, which in turn consist of triangles, and so on...'',
in the words of Hagedorn. Still later, shortly after Hagedorn's proposal,
the French mathematician Benoit Mandelbroit initiated the study
of such {\sl fractal behaviour} as a new field of mathematics.

\begin{figure}[htb]
\hfill{\epsfig{file=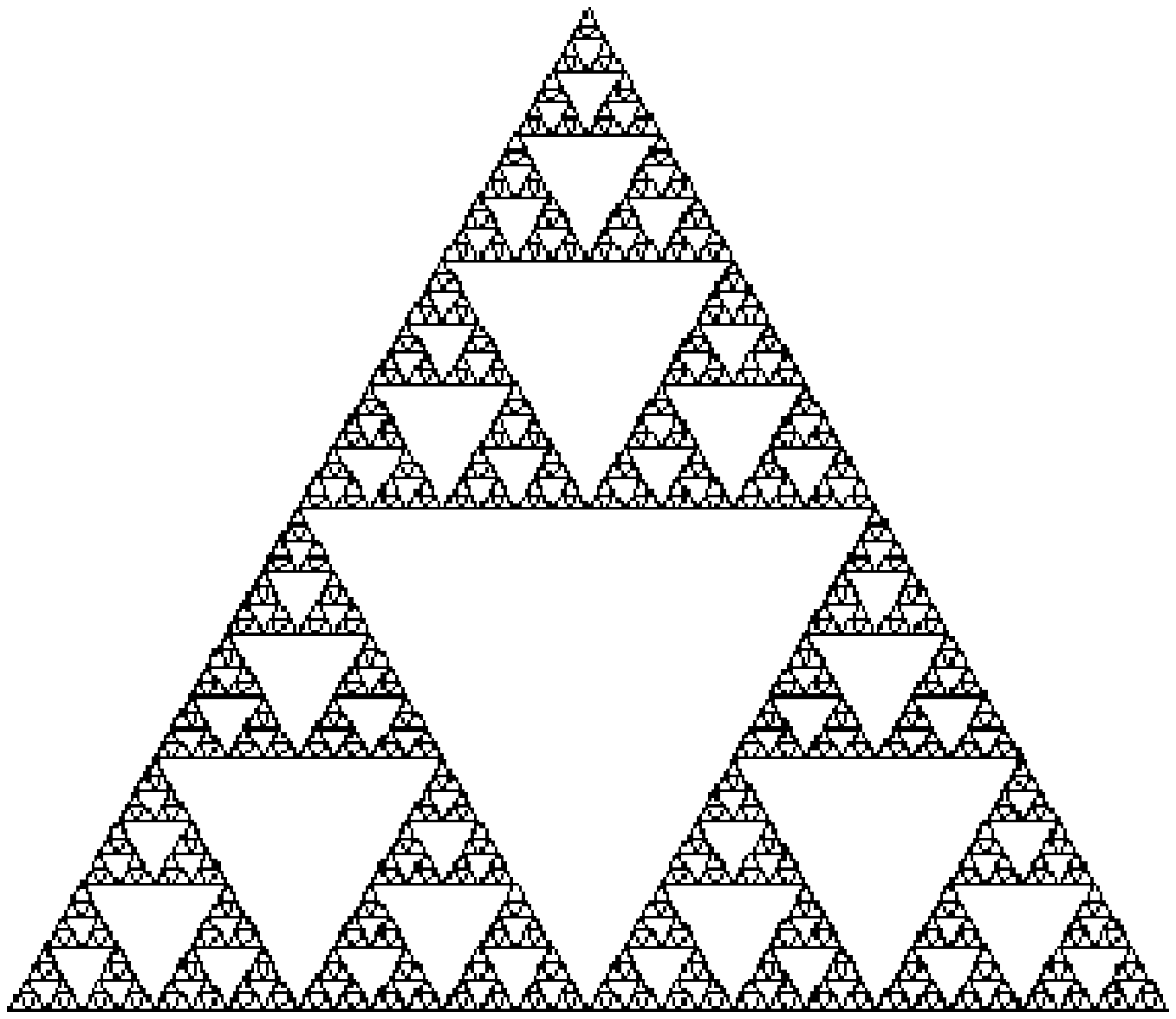,width=5cm,angle=-90}~
~~~~~~~~~~~~~~~~~~~~~~~~}
\end{figure}

\vskip-3.5cm
~~~~~~~~~~~~~~~~~~~The Sierpinski Triangle

\vskip3cm

Hagedorn had recalled a similar problem in number theory: how many ways are
there of decomposing an integer into integers? This was something already
adressed in 1753 by Leonhard Euler, and more than a century later by
the mathematician E.\ Schr\"oder in Germany. Finally G.\ H.\ Hardy and
S.\ Ramanujan in England provided an asymptotic solution. Let us here,
however, consider a simplified, easily solvable version of the problem
\cite{Blanchard}, in
which we count all possible different ordered arrangements $p(n)$ of an
integer $n$. So we have

\medskip

1=1 \hfill $p(1)=1=2^{n-1}$

2=2 ,1+1 \hfill $p(2)=2 = 2^{n-1}$

3= 3, 2+1, 1+2, 1+1+1 \hfill $p(3) = 4 = 2^{n-1}$

4= 4, 3+1, 1+3, 2+2, 2+1+1, 1+2+1, 1+1+2, 1+1+1+1 \hfill $p(4)=8 = 2^{n-1}$

\medskip

and so on. In other words, there are

\be
p(n) = 2^{n-1} = {1 \over 2}~ e^{n \ln 2}
\label{1}
\ee

ways of partitioning an integer $n$ into ordered partitions: $p(n)$ grows
exponentially in $n$. In this particular case, the solution could be found
simply by induction. But there is another way of getting it, more in line
with Hagedorn's thinking: ``large integers consist of smaller integers,
which in turn consist of still smaller integers, and so on...''
This can be formulated as an equation,
\be
\rho(n) = \delta(n\!-\!1) + \sum_{k=2}^n~ {1 \over k!}~ \prod_{i=1}^k
~\rho(n_i)~ \delta(\Sigma_i n_i\! -\!n).
\label{2}
\ee
It is quite evident here that the form of the partition number $\rho(n)$
is determined by a convolution of many similar partitions of smaller $n$.
The solution of the equation is in fact just the number of partitions of
$n$ that we had obtained above,
\be
\rho(n)=  z~p(n)
\label{3}
\ee
up to a normalization constant of order unity (for the present case, it
turns out that $z \simeq 1.25$). For Hagedorn, eq.\ (\ref{2}) expressed
the idea that the structure of $\rho(n)$ was determined by the structure
of $\rho(n)$ -- we now call this self-similar. He instead thought of the
legendary Baron von M\"unchhausen, who had extracted himself from a swamp
by pulling on his own bootstraps. So for him, eq.\ (\ref{2}) became a
{\sl bootstrap equation}.

\medskip

The problem Hagedorn had in mind was, of course, considerably more complex.
His heavy resonance was not simply a sum of lighter ones at rest, but it
was a system of lighter resonances in motion, with the requirement that the
total energy of this system added up to the mass of the heavy one. And
similarly, the masses of the lighter ones were the result of still ligher
ones in motion. The bootstrap equation for such a situation becomes
\be
\rho(m,V_0) = \delta(m\!-\!m_0) ~+
\sum_N {1\over N!} \left[ {V_0 \over (2\pi)^3} \right]^{N-1}
\hskip-0.2cm \int \prod_{i=1}^N ~[dm_i~ \rho(m_i)~ d^3p_i]
~\delta^4(\Sigma_i p_i - p),
\label{4}
\ee
where the first term corresponds to the case of just one lightest possible
particle, a ``pion''. The factor $V_0$, the so-called composition volume,
specified the size of the overall system, an intrinsic fireball size.
Since the mass of any resonance in the composition chain is thus determined
by the sum over phase spaces containing lighter ones, whose mass is
specified in the same way, Hagedorn called this form of bootstrap
``statistical''.

\medskip

After a number of numerical attempts by others, W.\ Nahm \cite{Nahm}
solved the statistical
bootstrap equation analytically, obtaining
\be
\rho(m,V_0) = {\rm const.}~m^{-3} \exp\{m/T_H\}.
\label{5}
\ee
So even though the partitioning now was not just additive in masses,
but included the kinetic energy of the moving constituents, the increase
was again exponential in mass. The coefficient of the increase,
$T_H^{-1}$, is determined by the equation
\be
{V_0 T_H^3 \over 2 \pi^2} (m_0/T_H)^2 K_2(m_0/T_H) = 2 \ln 2 - 1,
\label{6}
\ee
in terms of two parameters $V_0$ and $m_0$. Hagedorn assumed that
the composition volume $V_0$, specifying the intrinsic range of
strong interactions, was determined by the inverse pion mass as
scale, $V_0 \simeq (4 \pi/3) m_{\pi}^{-3}$. This leads to a scale factor
$T_H \simeq 150$ MeV. It should be emphasized, however, that this is
just one possible way to proceed. In the limit $m_0 \to 0$, eq.\ (\ref{6})
gives
\be
T_H = [\pi^2 (2\ln 2 -1)]^{1/3}~ V_0^{-1/3} \simeq 1/r_h,
\label{7}
\ee
where $V_0 = (4\pi /3) r_h^3$ and $r_h$ denotes the range of
strong interactions. With $r_h \simeq 1$ fm, we thus have
$T_H \simeq$ 200 MeV. From this it is evident that the exponential
increase persists also in the chiral limit $m_{\pi} \to 0$ and is in fact
only weakly dependent on $m_0$, provided the strong interaction
scale $V_0$ is kept fixed.

\medskip

The weights $\rho(m)$ determine the composition as well as the decay of
``resonances'', of fireballs. The basis of the entire formalism, the
 self-similarity postulate -- here in the form of the statistical bootstrap
condition -- results in an unending sequence of ever-heavier fireballs and in
an exponentially growing number of different states of a given mass $m$.

\medskip

Before we turn to the implications of such a pattern in thermodynamics,
we note that not long after Hagedorn's seminal paper, it was found that
a rather different approach, the dual resonance model \cite{DR1,DR2,DR3}
led to very much the
same exponential increase in the number of states. In this model, any
scattering amplitude, from an initial two to a final n hadrons, was assumed
to be determined by the resonance poles in the different kinematic channels.
This resulted structurally again in a partition problem of the same type, and
again the solution was that the number of possible resonance states of mass
$m$ must grow exponentially in $m$, with an inverse scale factor of the
same size as obtained above, some 200 MeV. Needless to say, this unexpected
support from the forefront of theoretical hadron dynamics considerably
enhanced the interest in Hagedorn's work.

{\large{\section{The Limiting Temperature of Hadronic Matter}}}

Consider a relativistic ideal gas of identical neutral scalar particles of
mass $m_0$ contained in a box of volume $V$, assuming Boltzmann statistics.
The grand canonical partition function of this system is given by

\be
{\cal Z}(T,V) = \sum_N {1 \over N!} \left[ {V \over (2\pi)^3} \int d^3p~
\exp\{-\sqrt{p^2+m_0^2}~/T\} \right]^N,
\ee

\noindent
leading to

\be
\ln {\cal Z}(T,V) = {VTm_0^2 \over 2\pi^2}~ K_2({m_0\over T}).
\ee

\medskip

\noindent
For temperatures $T\gg m_0$, the energy density of the system becomes

\be
\e(T) = - {1 \over V}~ {\partial~\!\ln~\!{\cal Z}(T,V) \over \partial~\!(1/T)}
\simeq {3 \over \pi^2}~ T^4,
\ee

\medskip

\noindent
the particle density

\be
n(T) =  {\partial~\!\ln~\!{\cal Z}(T,V) \over \partial~\!V}
\simeq {1 \over \pi^2}~ T^3,
\ee

\bigskip

\noindent
and so the average energy per particle is given by

\be
\omega \simeq 3~T.
\ee
\noindent
The important feature to learn from these relations is that, in the case
of an ideal gas of one species of elementary particles,
an increase of the energy of the system has three consequences:
it leads to

\vskip0.5cm
\begin{itemize}
\vspace*{-0.3cm}
\item{a higher temperature,}
\vspace*{-0.3cm}
\item{more constituents, and}
\vspace*{-0.3cm}
\item{more energetic constituents.}
\vspace*{-0.3cm}
\end{itemize}

\medskip

If we now consider an {\sl interacting} gas of such basic
hadrons and postulate that the essential form of the interaction
is resonance formation, then we can approximate the interacting medium as a
non-interacting gas of all possible resonance species \cite{B-U,DMB}.
The partition function of this resonance gas is

\be
\ln {\cal Z}(T,V) = \sum_i
{VTm_i^2 \over 2\pi^2}~\rho(m_i)~ K_2({m_i\over T})
\label{resgas}
\ee

\medskip

where the sum begins with the stable ground state $m_0$ and
then includes the possible resonances $m_i, i=1,2,...$ with
weights $\rho(m_i)$ relative to $m_0$.
Clearly the crucial question here is how to specify $\rho(m_i)$,
how many states there are of mass $m_i$. It is only at this point
that hadron dynamics enters, and it is here that Hagedorn introduced
the result obtained in his statistical bootstrap model.

\medskip

As we had seen above in eq.\ (\ref{5}), the density of states then
increases exponentially in $m$, with a coefficient $T_H^{-1}$ determined
by eq.\ (\ref{6}) in terms of two parameters $V_0$ and $m_0$.
If we replace the sum in the resonance gas partition function
(\ref{resgas}) by an integral and insert the exponentially growing mass
spectrum (\ref{5}), eq.\ (\ref{resgas}) becomes
$$
\ln {\cal Z}(T,V) \simeq
{V T \over 2\pi^2} \int dm~m^2 \rho(m_i)~ K_2({m_i\over T})
$$
\be
\sim V\left[{T \over 2\pi}\right]^{3/2} \int dm~m^{-{3/2}}
\exp\{-m \left[{1\over T} - {1\over T_H}\right]\}.
\label{div}
\ee

\medskip

\noindent
Evidently, the result is divergent for all $T > T_H$: in other words,
$T_H$ is the highest posssible temperature of hadronic matter. Moreover,
if we compare such a system with the ideal gas of only basic particles
( a ``pion'' gas), we find

\medskip

\hspace*{2.9cm}
pion gas \hskip 6cm resonance gas

\medskip

\hskip 3cm
$ n_{\pi} \sim  \e^{3/4} \hskip 6cm n_{res} \sim \e$

\hskip 3cm
$ \omega_{\pi} \sim \e^{1/4} \hskip 6cm \omega_{res} \sim {\rm const.}$

\medskip

\noindent
Here $n$ denotes the average number density of constituents, $\omega$ the
average energy of a constituent. In contrast to to the pion gas,
an increase of energy now leads to

\begin{itemize}
\vspace*{-0.2cm}
\item{a fixed temperature limit, $T \to T_H$,}
\vspace*{-0.3cm}
\item{the momenta of the constituents do not continue to increase, and}
\vspace*{-0.3cm}
\item{more and more species of ever heavier particles appear.}
\vspace*{-0.2cm}
\end{itemize}
We thus obtain a new, non-kinetic way to use energy, increasing the
number of species and their masses, not the momentum per particle.
Temperature is a measure of the momentum of the constituents, and if
that cannot continue to increase, there is a highest possible, a ``limiting''
temperature for hadronic systems.

\medskip

Hagedorn originally interpreted $T_H$ as the ultimate
temperature of strongly interacting matter. It is clear today that $T_H$
signals the transition from hadronic matter to a quark-gluon plasma.
Hadron physics alone can only specify its inherent limit; to go beyond
this limit, we need more information: we need QCD.

\medskip

As seen in eq.\ (\ref{5}), the solution of the statistical bootstrap
equation has the general form
\be
\rho(m,V_0) \sim m^{-a} \exp\{m/T_H\},
\label{boot}
\ee
with some constant $a$; the exact solution of eq.\ (\ref{4}) by Nahm gave
$a=3$. It is possible, however, to consider variations of the bootstrap
model which lead to different $a$, but always retain the exponential
increase in $m$. While the exponential form makes $T_H$ the upper limit
of permissible temperatures, the power law coefficient $a$ determines
the behavior of the system at $T=T_H$. For $a=3$, the partition function
(\ref{div}) itself exists at that point, while the energy density as first
derivative in temperature diverges there. This is what made Hagedorn
conclude that $T_H$ is indeed the highest possible temperature of matter:
it would require an infinite energy to reach it.

\medskip

Only a few years later it was, however, pointed out by N.\ Cabibbo and
G.\ Parisi \cite{C-P} that larger $a$ shifted the divergence at $T=T_H$ to
ever higher derivatives. In particular, for $4> a > 3$, the energy
density would remain finite at that point, shifting the divergence
to the specific heat as next higher derivative. Such critical behavior was
in fact
quite conventional in thermodynamics: it signalled a phase transition leading
to the onset of a new state of matter. By that time, the quark model and
quantum chromodynamics as fundamental theory of strong interactions
had appeared and suggested the existence of a quark-gluon plasma as the
relevant state of matter at extreme temperature or density. It was
therefore natural to interpret the Hagedorn temperature $T_H$ as the
critical transition temperature from hadronic matter to such a plasma. This
interpretation is moreover corroborated by a calculation of the
critical exponents \cite{HScrit} governing the singular behavior of the
resonance gas thermodynamics based on a spectrum of the form (\ref{boot}).

\medskip

It should be noted, however, that in some sense $T_H$ did remain the
highest possible temperature of matter as we know it. Our matter exists in
the physical vacuum and is constructed out of fundamental building blocks
which in turn have an independent existence in this vacuum. Our matter
ultimately consists of and can be broken up into nucleons; we can
isolate and study a single
nucleon.
The quark-gluon plasma, on the other hand, has its own ground state, distinct
from the physical vacuum, and its constituents can exist only in a dense
medium of other quarks -- we can never isolate and study a single quark.

\medskip

\begin{figure}[h]
\centerline{\psfig{file=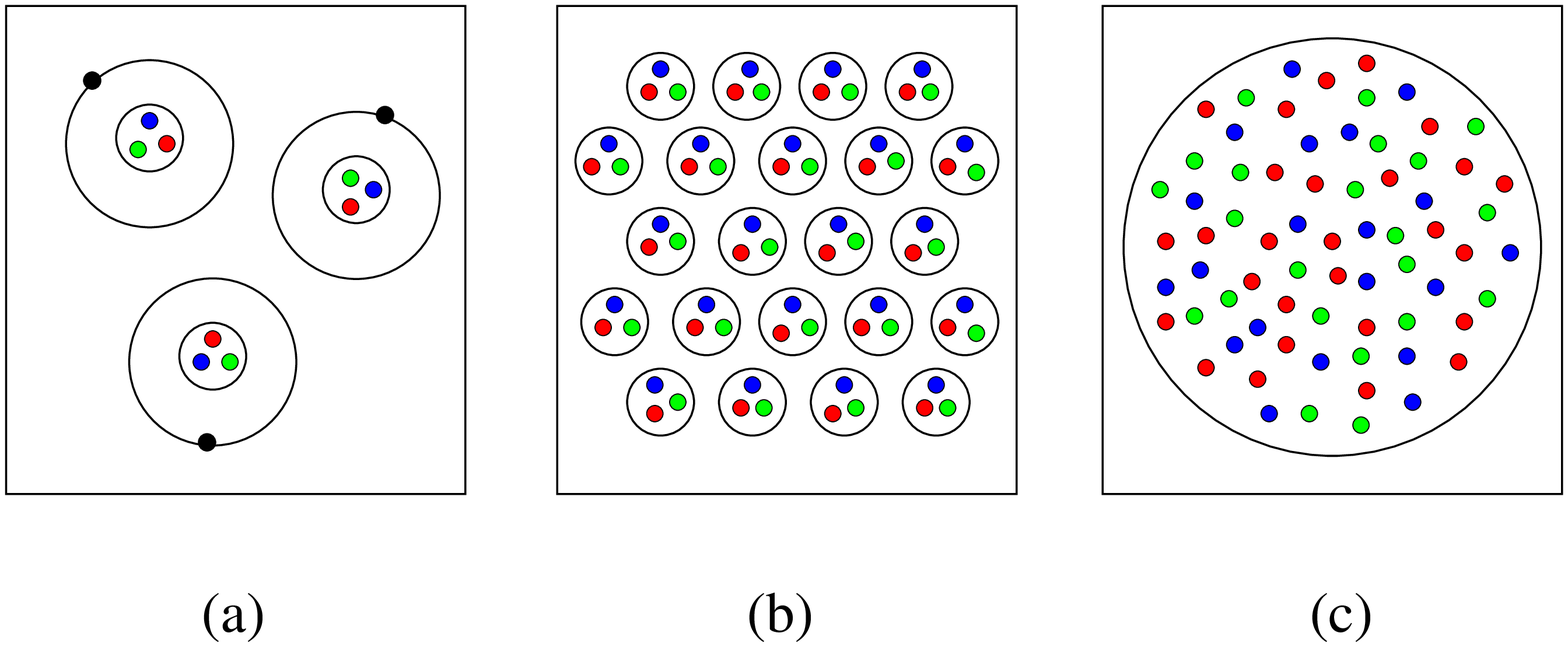,width=11cm}}
\caption{Schematic view of matter for increasing density, from atomic (a) to
nuclear (b) and then to quark matter (c).}
\label{states}
\end{figure}

That does not mean, however, that quarks are eternally confined to a given
part of space. Let us start with atomic matter and compress that to form
nuclear matter, as it exists in heavy nuclei. At this stage, we have nucleons
existing in the physical vacuum. Each nucleon consists of three quarks, and
they are confined to remain close to each other; there is no way to break
up a given nucleon into its quark constituents. But if we now continue to
compress, then eventually the nucleons will then penetrate each other,
until we reach a dense medium of quarks. Now each quark finds in its
immediate neighborhood
many other quarks besides those which were with it in the nucleon stage.
It is therefore no longer possible to partition quarks into nucleons;
the medium consists of unbound quarks, whose interaction becomes ever weaker
with increasing density, approaching the limit of asymptotic freedom
predicted by QCD. Any quark can now move freely throughout the medium:
we have quark liberation through swarm formation. Wherever a quark
goes, there are many other quarks nearby. The transition from
atomic to quark matter is schematically illustrated in Fig.\ \ref{states}.

\medskip

We have here considered quark matter formation through the compression of
cold nuclear matter. A similar effect is obtained if we heat a meson gas;
with increasing temperature, collisions and pair production lead to an
ever denser medium of mesons. And according to Hagedorn, also of ever
heavier mesons of an increasing degeneracy. For Hagedorn, the fireballs
where pointlike, so that the overlap we had just noted simply does not
occur. In the real world, however, they do have hadronic size, so that
they will in fact interpenetrate and overlap before the divergence of
the Hagedorn resonance gas occurs \cite{Zinov}. Hence now again there
will be a transition from resonance gas to a quark-gluon plasma, now formed
by the liberation of the quarks and gluons making up the resonances.

\medskip

At this point, it seems worthwhile to note an even earlier approach leading
to a limiting temperature for hadronic matter. More than a decade before
Hagedorn, I.\ Ya.\ Pomeranchuk \cite{Pom} had pointed out that a crucial
feature of hadrons is their size, and hence the density of any hadronic
medium is limited by volume restriction: each hadron must have its own volume
to exist, and once the density reaches the dense packing limit, it's the end
for hadronic matter. This simply led to a temperature limit, and for an ideal
gas of pions of 1 fm radius, the resulting temperature was again around
200 MeV. Nevertheless, these early results remained largely unnoticed until
the work of Hagedorn.

\medskip

Such geometric considerations do, however, lead even further. If hadrons are
allowed to interpenetrate, to overlap, then percolation theory predicts
two different states of matter \cite{Perc1,Perc2}: hadronic matter, consisting
of isolated hadrons or finite hadronic clusters, and a medium formed as an
infinite sized cluster of overlapping hadrons. The transition from one to
the other now becomes a genuine critical phenomenon, occurring at a critical
value of the hadron density.

\medskip

We thus conclude that the pioneering work of Rolf Hagedorn opened up the
field of critical behavior in strong interaction physics, a field in which
still today much is determined by his ideas. On a more theoretical level,
the continuation of such studies was provided by finite temperature lattice
QCD, and on the more experimental side, by resonance gas analyses of
the hadron abundances in high energy collisions. In both cases, it was
found that the observed behavior was essentially that predicted by
Hagedorn's ideas.

{\large{\section{Resonance gas and QCD thermodynamics}}}

With the formulation of Quantum Chromodynamics (QCD) as a
theoretical framework for the strong interaction force among
elementary particles it became clear that the appearance of the ultimate
Hagedorn temperature $T_H$ signals indeed  the transition from hadronic phase
to a new phase of strongly interacting matter, the quark-gluon plasma
(QGP) \cite{HS}.
As QCD is an asymptotically free theory, the interaction between quarks and
gluons vanishes
logarithmically with increasing temperature, thus at very high
temperatures the QGP  effectively
behaves like an ideal gas of quarks and gluons.

\medskip

Today we have detailed information, obtained from numerical calculations
in the framework of finite temperature lattice Quantum Chromodynamics
\cite{first1,first2,first3,review,fodor}, about the thermodynamics of 
hot and dense
matter. We know the transition temperature to the QGP and the temperature
dependence of basic bulk thermodynamic observables such as the
energy density and the pressure \cite{fk,fodor}. We also begin to have results
on fluctuations and correlations of conserved charges
\cite{C1,C2,C3}.

\medskip

The recent increase in numerical accuracy  of lattice QCD calculations
and their extrapolation to the continuum limit make it possible  to
confront the fundamental results of QCD  with Hagedorn's
concepts \cite{rh2, Hagedorn3}, which provide a theoretical scenario
for the thermodynamics of strongly interacting
hadronic matter
\cite{C3,taw1,taw2,frit}.

\medskip

In particular, the
equation of state calculated  on the lattice at vanishing and finite
chemical potential, and restricted to the confined hadronic phase, can be
directly compared to that obtained from the partition function (\ref{div})
of the hadron resonance gas, using the form (\ref{boot}) introduced by
Hagedorn for a continuum mass spectrum.
Alternatively, as first approximation, one can also consider a discrete
mass spectrum which accounts for all experimentally known hadrons and
resonances. In this case the continuum partition function of the Hagedorn
model is expressed by Eq.\ (\ref{resgas}) with $\rho(m_i)$ replaced by the
spin
degeneracy factor of the $i^{th}$ hadron, and with the summation taken over
all  known resonance species listed by
the Particle Data Group \cite{pdg}.

\medskip

With the above assumption on the dynamics and the mass spectrum,
the Hagedorn resonance gas partition function \cite{rh2, Hagedorn3}
can be calculated exactly and expressed as a sum of
one--particle partition functions $Z^1_i$ of all hadrons and
resonances,

\be
 \ln Z(T,V)=\sum_i Z_i^1(T,V).
\label{qq1}
\ee
For particles of mass $m_i$ and  spin
degeneracy factor $g_i$,  the one--particle partition function
$Z^1_i$, in the Boltzmann approximation,  reads

\be  Z^1_i(T,V)= g_i{{VTm_i^2}\over {2\pi^2}} K_2({{m_i}\over T}). \label{eeq2} \ee
Due to the factorization of the partition function in
Eq.~(\ref{qq1}),  the energy density and  the   pressure of the Hagedorn
resonance gas  with a discrete  mass spectrum,
can also be expressed as a  sum over single particle contributions
\be
 \epsilon=\sum_i\epsilon_i^1~~,~~P=\sum_i P_i^1, \label{qq3}
 \ee
with
\begin{eqnarray}
{{\epsilon_i^1}\over {T^4}} &=& \frac{g_i}{2\pi^2}\;
(\frac{ m_i}{T})^3
\;\left[\frac{3\;K_2(\beta m_i)}{\beta m} +
  \;K_1(\beta m_i)\right]\label{eqq4}  \label{eqq5}
\end{eqnarray}
and
\begin{eqnarray}
\frac{P_i^1}{T^4} &=& \frac{g_i}{2
\pi^2}\;(\frac{m_i}{T})^3 \; K_2(\beta m_i) \label{eqq6},
\end{eqnarray}
where $\beta =1/T$ and   $K_1$ and $K_2$ are modified Bessel functions.
At vanishing chemical potentials and at finite temperature, the energy density
$\epsilon$, the entropy density $s$ and the pressure $P$,
are connected through the thermodynamic relation,
\begin{eqnarray}
\epsilon=-P+sT.\label{eq6}
\end{eqnarray}
Summing up in Eq.~(\ref{qq3})  the contributions from
experimentally known hadronic states, constitutes the resonance gas
\cite{rh2, Hagedorn3,redb}  for the
 thermodynamics of the hadronic  phase of QCD.
Taking into account contributions of all mesonic and baryonic resonances
with masses up to 1.8 GeV and 2.0 GeV, respectively, amounts to 1026
resonances.

\medskip

\begin{figure}[!t]
 \center\includegraphics[width=3.50in,height=2.3in]{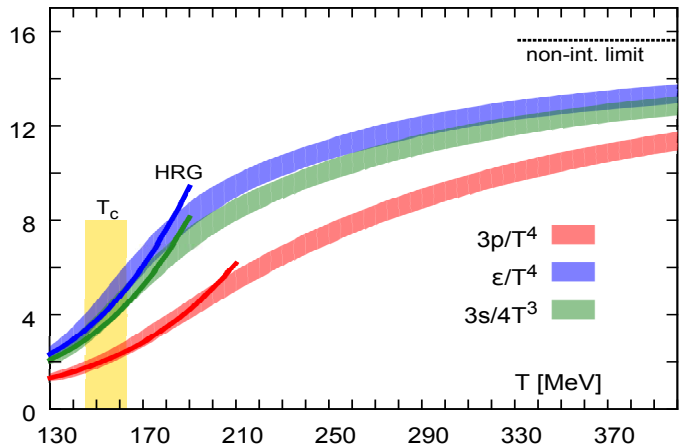}\hskip 0.9cm
 \vskip -0.1cm   \caption{The
 normalized pressure $P(T)$, the energy density $\epsilon(T)$  and the entropy
 density  $s(T)$ obtained in lattice QCD calculations  as a function of
 temperature. The dark lines show the prediction of the Hagedorn resonance gas
 for discrete mass spectrum, eqs.~(\ref{qq3}) - (\ref{eq6}).  The
 lattice results are  from Ref.\ \cite{fk}.}\label{fig1}
 \end{figure}

\medskip

The crucial question thus is if the equation of state
introduced by Hagedorn can describe the
corresponding results obtained from  QCD within lattice approach. In
Fig.~\ref{fig1} we show the temperature dependence of the energy
density, pressure and the entropy density obtained recently in
lattice QCD studies with  physical  masses of up, down and strange
quarks \cite{fk}.  The bands in lattice QCD results indicate error bars due
to extrapolation to the continuum limit.
The vertical band marks the temperature,
$T_c = (154 \pm 9)$ MeV,  which  within error,
is the crossover temperature from a hadronic phase to a quark-gluon
plasma \cite{tc}.
These QCD results are compared in Fig.\ \ref{fig1} to
Hagedorn's resonance gas model (HRG)
formulated for a  discrete mass spectrum in
  Eqs.~(\ref{qq3}) and  (\ref{eq6}).
There is excellent agreement between the Hagedorn model results for the
equation of states and the corresponding lattice data.
All bulk thermodynamical observables  are very strongly
changing with temperature when approaching the  deconfinement
transition temperature. In Hagedorn's formulation, this behavior is
well understood in terms of increasing resonance contributions.
Although the HRG formulated for  discrete mass spectrum   does not
exhibit any critical
behavior, it nevertheless reproduces remarkably well the lattice
results in the hadronic phase. This agreement has now been extended to
an analysis of
fluctuations and correlations of conserved charges as well.

\medskip

The excellent description of the lattice QCD results by
Hagedorn's model
justifies the claim {\sl that resonances are indeed the
essential  degrees of freedom near deconfinement}. Thus,
on the thermodynamical level, modeling hadronic interactions
by formation and excitation of resonances, as introduced by Hagedorn,
is an excellent approximation of strong interactions.

\bigskip\bigskip

{\large{\section{Resonance Gas and Heavy Ion Collisions}}}

Long before  lattice QCD could  provide a  direct  evidence that strong
interaction thermodynamics  can be   quantified  by the resonance
gas partition function, Hagedorn's  concept was verified phenomenologically
by considering particle production in elementary and heavy ion collisions
\cite{jc,redf,heppe,beca,redb}.
In a strongly interacting medium, one includes the conservation of electric
charge, baryon number and strangeness.  In this case, the partition function
of Hagedorn's thermal model  depends not only on temperature but also on
chemical
potential $\vec\mu$, which guarantees,  that charges are conserved on an
average.
For a non vanishing $\vec\mu$, the   partition function  Eq. (\ref{qq1})
is replaced by
\be
 \ln Z(T,V,\vec\mu )=\sum_i  Z_i^1(T,V,\vec\mu),
\label{par}
\ee
with $\vec \mu =(\mu_B,\mu_S,\mu_Q)$,   where $\mu_i$ are the chemical
potentials
related to the baryon number, strangeness and  electric charge conservation,
respectively.

\medskip

For particle $i$ carrying  stran\-geness $S_i$, the  baryon number $B_i$, the
electric
charge $Q_i$ and the spin--isospin degeneracy factor $g_i$, the one particle
partition function, reads
 \be
Z_i^1(T,V,\vec\mu) =\frac{ Vg_iTm_i^2 }{ 2\pi^2 }
K_2({{m_i}/T}) \exp \left(
\frac{B_i\mu_B+S_i\mu_S+Q_i\mu_Q}{T} \right).
 \label{parp}
\ee
For $\vec\mu=0$ one recovers the result  from  Eq. (\ref{eeq2}).

\medskip

The calculation of a density $n_i$ of  particle $i$  from the partition
function Eq. (\ref{par}   )
is rather straightforward \cite{hagedornred}. It amounts to the replacement
$Z_i^1\to \gamma_i Z_i^1$ in Eq. (\ref{par})   and taking a derivative  with
respect
to
the particle fugacity $\gamma_i$, as
\be
n_i=\frac{ \langle N_i\rangle^\mathrm{th} }{V}=\left.\frac{\partial \ln Z}
{\partial \gamma_i}\right|_{\gamma_i=1},\label{denj}
\ee
consequently,  $n_i=Z_i^1/V$ with  $Z_i^1$ as in  Eq.  (\ref{parp}).

\medskip

The   Hagedorn  model, formulated in Eq.  (\ref{par}),  describes  bulk
thermodynamic properties
and  particle composition of a  thermal fireball at finite temperature  and at
non
vanishing charge densities. If such a fireball is created  in high energy
heavy
ion collisions, then yields of different hadron species  are fully quantified
 by thermal parameters.  However,  following   Hagedorn's idea,  the
 contribution
of resonances   decaying   into lighter particles,   must be included~
\cite{rh2,Hagedorn3}.

\medskip

In  Hagedorn's thermal model, the average number $\langle N_i\rangle$
of particles $i$   in volume $V$ and at temperature $T$ that carries
strangeness
$S_i$, the baryon number $B_i$, and the electric charge $Q_i$, is  obtained
from
Eq. (\ref{par}),  see~\cite{rh2,Hagedorn3}
\be
 \langle N_i^{}\rangle (T,\vec\mu)~=~\langle N_i\rangle^\mathrm{th} (T,\vec\mu)
 +{\sum}_j\Gamma_{j\to i}
\langle N_j\rangle^{th,R}(T,\vec\mu)\label{denn} .
\ee
The first term  in Eq. (\ref{denn})  describes the thermal average number of
particles
of species $i$ from Eq. (\ref{denj})   and the second term describes overall
contribution
from  resonances. This term is taken as a sum of all resonances that decay
into
particle $i$. The $\Gamma_{j\to i}$ is the corresponding decay branching ratio
of
$j\to i$. The  multiplicities of resonances $\langle N_j\rangle^{th,R}$  in
Eq. (\ref{denn}), are obtained from Eq. (\ref{denj}).

\medskip

\begin{figure}
\centerline{\psfig{file=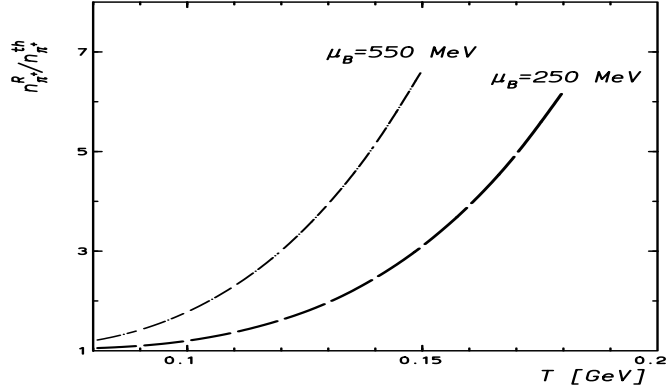, width=3.50in, height=2.9in, angle=180}}
\vspace*{-1.3cm}
\caption{The ratio of the total density of  positively charged pions,
$n_{\pi^+}^R$ from  Eq. (\ref{denn}),  and the  density of thermal pions,
$n_{\pi^+}^\mathrm{th}$ from Eq. (\ref{denj}). The calculations are done in the
Hagedorn
resonance gas model for  $\mu_B=$250 MeV and $\mu_B=$550 MeV at different
temperatures.   }\label{fig2}
 \end{figure}

The importance of resonance contributions to the total particle yield in
Eq. (\ref{denn})
is illustrated in Fig. (\ref{fig2})  for charge pions. In Fig. (\ref{fig2})
we show  the
ratio of the
total number of charge pions   from Eq. (\ref{denn})   and the number of
prompt
pions  from
Eq. (\ref{denj}). The  ratio is  strongly increasing with temperature and
chemical
  potential.
This is  due to an increasing contribution of mesonic and baryonic
resonances. From
Eq. (\ref{fig2})  it is clear,  that at high temperature and/or   density,
the
overall
multiplicity of pions  is  mostly due to   resonance decays.

\medskip

The particle yields in Hagedorn's model Eq. (\ref{denn}) depend, in general,
on five parameters. However, in high energy heavy ion collisions, only three
parameters are independent. The isospin asymmetry,  in the initial state fixes
the charge chemical potential and the strangeness neutrality condition
eliminates
the strange chemical potential. Thus, on the level of particle multiplicity, we
are
left with temperature $T$ and the baryon chemical potential $\mu_B$ as
independent
parameters, as well as, with fireball volume as an overall normalization
factor.

\medskip

Hagedorn's thermal model introduced  in Eq. (\ref{denn})  was successfully
applied   to describe particle yields measured in heavy ion collisions.
The model was compared with available experimental data obtained in a broad
energy range from AGS up to LHC. Hadron multiplicities ranging from pions
to omega baryons and their ratios, as well as composite objects like e.g.
deuteron or alpha particle, were used to verify if there was  a set of
thermal parameters $(T,\mu_B)$ and $V$,  which simultaneously reproduces
all measured yields.

\medskip

The systematic studies  of particle production extended  over more than
two  decades, using experimental results at different beam energies,
have revealed a clear  justification, that in central heavy ion collisions
particle yields are indeed consistent with the expectation of the Hagedorn
thermal  model. There is also a clear pattern of the energy,
$\sqrt s$-dependence of thermal parameters. The temperature is
increasing with $\sqrt s$,  and at the SPS energy essentially saturates
at the value,  which corresponds to the transition temperature from a
hadronic phase to a QGP, as obtained in LQCD. The chemical potential,
on the other hand, is gradually decreasing with $\sqrt s$ and almost
vanishes at the LHC.

\medskip

\begin{figure}
\centerline{\psfig{file=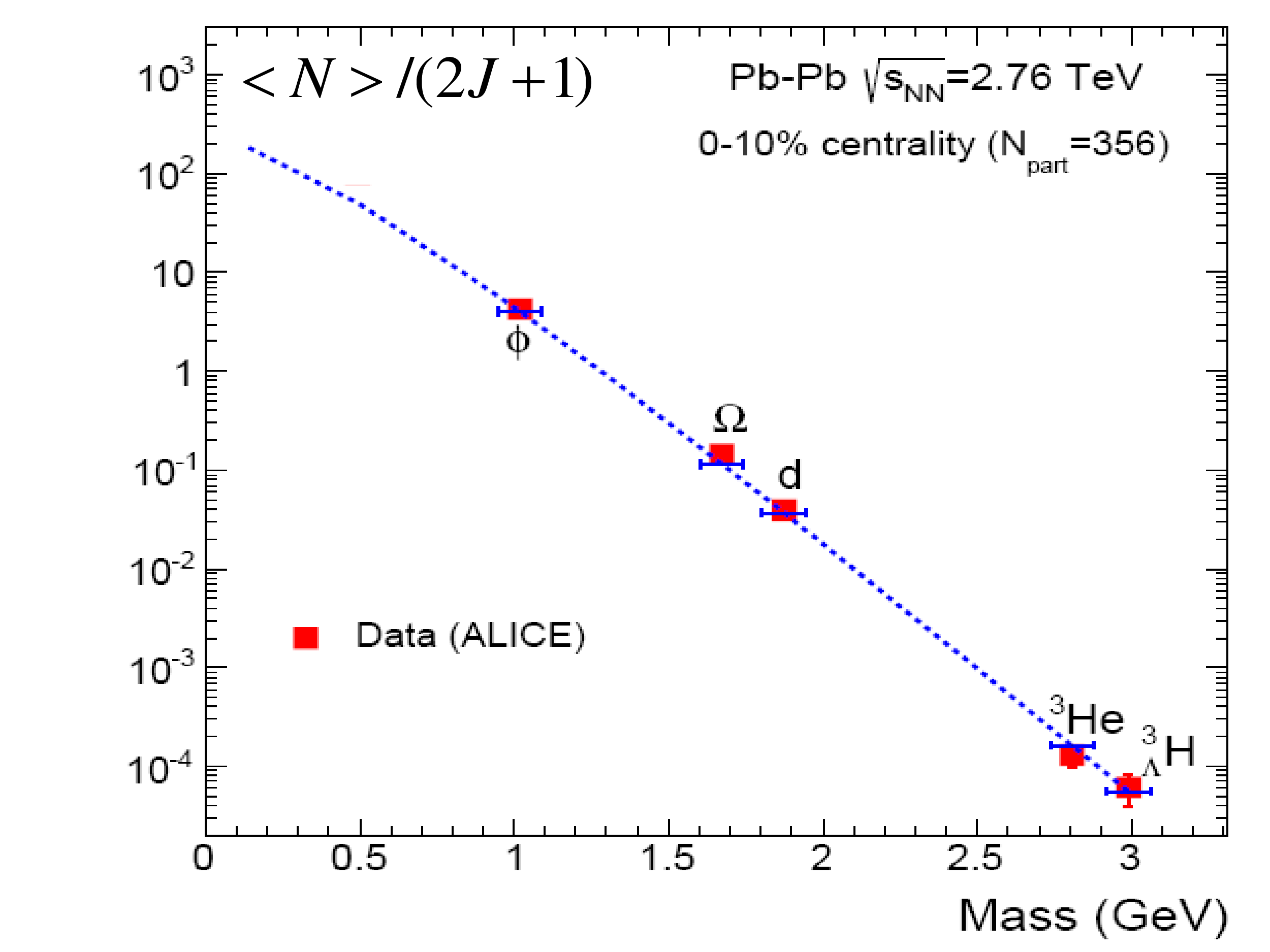, width=3.50in, height=2.2in}}
\caption{Yields of several different particle species per unit rapidity
normalized to spin degeneracy factor as a function of their mass.
Data are from ALICE collaboration taken at the LHC in central Pb-Pb
collisions. The line is the  Hagedorn  thermal  model result, Eq.
(\ref{raf}), see Ref.\cite{andronic1,andronic2}.}\label{fig3}
 \end{figure}

In Fig. (\ref{fig3})  we show, as an illustration, a  comparison of
Hagedorn's thermal model   and  recent data on selected  particle yields,
obtained by  ALICE collaboration  in central Pb-Pb collisions at midrapidity
at the LHC energy~\cite{andronic1,andronic2}. At such high collision  energy,
particle yields from Eq. (\ref{denn})  are  quantified entirely  by the
temperature and the fireball volume.\footnote{The chemical potential
$\vec\mu$ in Eq. (\ref{denn})  vanishes, since at the LHC   and at
midrapidity particles and their antiparticles are produced symmetrically.}
Thus, there is transparent  prediction  of  Hagedorn's model
Eq. (\ref{denn}), that yields of heavier particles $\langle N_{i}\rangle$
with no resonance decay contributions, normalized to their spin degeneracy
factor $g_i=(2J+1)$,    should be quantified by
\be
\frac{ \langle N_{i}\rangle}{ 2J+1 }\simeq VT^3 \left(\frac{{m_i}}{ 2\pi T}
\right)^{3/2}
 \exp( -{{m_i}/ T}),
\label{raf}
\ee
where we have used Eq. (\ref{denj})  and the asymptotic expansion of the
Bessel function, $K_2(x)\sim x^{-1/2}\exp(-x)$, valid for large $x$.

\medskip

In Fig. (\ref{fig3})  we show the yields of particles with no resonance
contribution, like  $\phi$, $\Omega$, the deuteron `d',  $^3$He and the
hypertriton  $_\Lambda^3$He,  normalized to their  spin degeneracy factor,
as a function of particle  mass. Also shown in this figure is the prediction
from Eq. (\ref{raf})  at $T\simeq 156$  and  for   volume
$V\simeq 5000$ fm$^3$~\cite{andronic1,andronic2}. There is a clear coincidence
of  data taken in Pb-Pb collisions at the LHC and  predictions of the
Hagedorn model Eq.  (\ref{raf}).  Particles with no resonance contribution
measured by ALICE collaboration  follow  the Hagedorn's expectations that
they  are produced from a thermal fireball at common   temperature.
A similar agreement of Hagedorn's  thermal  concept  and experimental
data taken in central heavy ion collisions has been found for different
yields of  measured particles and collision energies from AGS, SPS, RHIC
and  LHC~\cite{redb}.

\bigskip\bigskip

{\large{\section{Particle Yields and Canonical Charge Conservation}}}

The Hagedorn thermodynamical model for particle production,
was originally applied  to quantify and understand particle
yields and spectra measured in elementary collisions --
there were no data available from heavy ion collisions.

\medskip

Initial work on particle production by Hagedorn begins in 1957
in collaboration with F. Cerulus when   they apply the Fermi
phase space model. In this
microcanonical approach, conservation laws  of  baryon number
or electric charge are implemented exactly. Almost  15 years
later the production of complex light antinuclei, such as  \medskip
anti-He$^3$,  preoccupied Hagedorn~\cite{rh2,Hagedorn3}.
He realized  and discussed clearly the need to find a path
to enforce exact conservation of baryon number to describe
the  anti-He$^3$ production correctly within the canonical
statistical formulation.

\medskip

Indeed, applying in \pp ~reactions the thermal model without
concern for conservation of baryon number  overestimates the
production  of  anti-He$^3$ in proton-proton collisions by
seven orders of magnitude \cite{rh2,rh4,Hagedorn3}. The reason
was that when the number of particles in the interaction volume
is small,  one has to take into account the fact that the
production of anti-He$^3$ must be  accompanied  by the production
of another three nucleons with energy $E_N$, in order  to  exactly
conserve the baryon number. Thus, in case that the production
of anti-He$^3$ is not  originating from reservoir of many
antiquarks or antinucleons already present in a large volume,
but  is rather originating  from some small volume $V_{pp}$
that  is present in pp~ collisions,  the abundance  of anti-He$^3$
will not be proportional to the single standard Boltzmann factor,
as in Eq. (\ref{raf})
\begin{equation}
n_{\overline{\,\mathrm{He}}^3} \sim
\exp \left( -{m_{\,\overline{\mathrm{He}}^3}/T} \right),
\end{equation}
but   is accompanied by additional Boltzmann factors that characterize
the production of the associated    nucleons,  needed in order to conserve
baryon number~\cite{rh2,Hagedorn3}
\begin{equation}
n_{\,\overline{\,\mathrm{He}}^3} \sim
\exp \left( -{m_{\overline{\,\mathrm{He}}^3}/ T} \right)
\left[ V_{\rm pp} \int \frac{d^3p}{ (2\pi)^3}
\exp \left( -\frac{E_N}{ T}\right)\right]^3.
\end{equation}
This   suppresses  the  rate  and  introduces  a   strong power-law
dependence on volume $V_{pp}$ for  the anti-He$^3$ yield.

\medskip

The problem of exact conservation of discrete quantum numbers in a
thermal model formulated in early '70s by Hagedorn in the context of
baryon number conservation remained unsolved for a decade. When the
heavy ion QGP research program was approaching and strangeness emerged
as a potential QGP signature, Hagedorn pointed out  the need to consider
exact conservation of strangeness  \cite{privat}.   This is the reason
that the old problem of baryon number conservation was solved in the new
context of strangeness conservation ~\cite{rafeldan,tounsi,oeschler}.
A more general solution, applicable to {\em all} discrete conserved
charges, abelian and non-abelian, was also introduced   in
Ref. ~\cite{turko1}   and expanded in ~\cite{turko2,petrov,rafm,
hagedornred,Derreth,Elze}.
Recently,   it  has become clear that a similar treatment should be
followed  not only  for  strangeness    but also for  charm abundance
study in high energy $e^+e^-$  collisions \cite{s1,b1}.

\medskip

To summarize this section, we note that the usual form of the
statistical model, based on a grand canonical formulation  of the
conservation laws, cannot be used when either the temperature or the
volume or both are small. As a rough estimate, one needs $V T^3 > 1$
for a grand canonical description to hold \cite{rafeldan,hagedornred}.
In the opposite limit, a path was found  within the canonical ensemble
to enforce  charge conservations exactly.

\medskip

The canonical approach has been   shown to provide   a consistent
description of particle production in  high energy  hadron-hadron,
$e^+e^-$  and peripheral heavy ion    collisions \cite{redb,rafm,s1,b1}.
In the context of developing strangeness as signature of QGP,  such a model also provides,  within
the realm of assumed strangeness chemical equilibrium,  a description
of  an observed increase of single-   and multi-strange particle  yields
from    \pp,~ \pA  ~to \hAA ~collisions  and its energy dependence \cite{tounsi}.

\bigskip\bigskip

{\large{\section{Concluding Remarks}}}

Rolf Hagedorn's work, introducing concepts from statistical mechanics
and from the mathematics of self-similarity into the analysis of high
energy multiparticle production, started a new field of research, alive
and active until today. On the theory side, the limiting temperature of
hadronic matter and the behavior of the Hagedorn resonance gas approaching
that limit were subsequently verified by first principle calculations
in finite temperature QCD. On the experimental side, particle yields as
well as, more recently, fluctuations of conserved quantities, were also
found to follow the pattern predicted by the Hagedorn resonance gas.
Rarely has an idea in physics risen from such humble and little
appreciated beginnings to such a striking vindication. So perhaps
it is appropriate to close with a poetic summary one of us (HS)
formulated some twenty years ago for a Hagedorn-Fest, with a slight update.

\vskip0.5cm


\begin{center}
\bigskip
{\bf Hot Hadronic Matter}

\medskip

{\sl (A Poetic Summary)}

\bigskip

In days of old

a tale was told

of hadrons ever fatter.

Behold, my friends, said Hagedorn,

the ultimate of matter.

\bigskip

Then Muster Mark

called in the quarks,

to hadrons they were mated.

Of colors three, and never free,

all to confinement fated.

\bigskip

But in dense matter,

their bonds can shatter

and they can freely move around.

Above $T_H$, their colors shine,

as the QGP is found.

\bigskip

Said Hagedorn,

when quarks were born

they had different advances.

Today they form, as we can see,

a gas of all their chances.
\end{center}

\vskip1cm

\centerline{\bf Acknowledgement:}

\medskip

K.R. acknowledges
 support  by the Polish Science Foundation (NCN), under
 Maestro grant DEC-2013/10/A/ST2/00106.



\begin{thebibliography}{99}




  \bibitem{rh2}
  R.~Hagedorn:
  ``Statistical thermodynamics of strong interactions at high-energies,''
  Nuovo Cim. Suppl. {\bf 3}, 147 (1965)

 \bibitem{rh3}
  R.~Hagedorn and J.~Ranft:
  ``Statistical thermodynamics of strong interactions at high-energies. 2.
Momentum spectra of particles produced in \pp ~collisions,''
  Nuovo Cim. Suppl. {\bf 6}, 169 (1968)

\bibitem{rh4}
R.~Hagedorn:
  ``Statistical thermodynamics of strong interactions at high energies. 3.
Heavy-pair (quark) production rates,''
  Nuovo Cim. Suppl.  {\bf 6}, 311  (1968)

\bibitem{Hagedorn2}
R.~Hagedorn:
  ``Hadronic matter near the boiling point,''
  Nuovo Cim. A {\bf 56}, 1027 (1968)

\bibitem{Hagedorn3}
R.~Hagedorn: {\sl Thermodynamics of Strong Interactions},
CERN-Report 71-12 (1971)


\bibitem{Blanchard} Ph.~Blanchard, S.~Fortunato and H.~Satz:
  ``The Hagedorn temperature and partition thermodynamics,''
Eur. Phys. J. C {\bf 34}, 361 (2004)

\bibitem{Nahm} W.~Nahm:    ``Analytical solution of the statistical
bootstrap model,''
 Phys. B {\bf 45}, 525 (1972)

\bibitem{DR1} G.~Veneziano:   ``Construction of a crossing - symmetric,
Regge behaved amplitude for linearly rising trajectories,''
Nuovo Cim. A {\bf 57}, 190 (1968)

\bibitem{DR2} K.~Bardakci and S.~Mandelstam:   ``Analytic solution of the
linear-trajectory bootstrap,''
Phys. Rev. {\bf 184}, 1640 (1969)

\bibitem{DR3} S.~Fubini and G.~Veneziano:   ``Level structure of
dual-resonance models,''
Nuovo Cim. A {\bf 64}, 811 (1969)

\bibitem{B-U} E.~Beth and G.E.~Uhlenbeck:  ``The quantum theory of
the non-ideal gas. II. Behaviour at low temperatures,''
Physica {\bf 4}, 915 (1937)

\bibitem{DMB} R.~Dashen, S.-K.~Ma and H.J.~Bernstein:  ``S Matrix
formulation of statistical mechanics,''
Phys. Rev. {\bf 187}, 345 (1969)

\bibitem{C-P} N.~Cabibbo and G.~Parisi:   ``Exponential hadronic
spectrum and quark liberation,''
Phys. Lett. B {\bf 59}, 67 (1975)

\bibitem{Hardy} G.H.~Hardy and S.~Ramanujan:``Asymptotic formulae
in combinatory analysis''
Proc.London  Math. Soc. {\bf 17}, 75 (1918)

\bibitem{HScrit} H.~Satz:   ``Critical behavior of hadronic matter.
1. Critical point exponents,''
Phys. Rev. {\bf D19}, 1912 (1979)

\bibitem{Zinov} M.~I.~Gorenstein, V.~A.~Miransky, V.~P.~Shelest,
G.~M.~Zinovev and H.~Satz:
  ``The physical content of the statistical bootstrap,''
 Nucl. Phys. {\bf B76}, 453 (1974)

\bibitem{Pom} I.Ya.~Pomeranchuk:   ``On the theory of multiple
particle production in a single collision,''
Dokl. Akad. Nauk {\bf 78}, 889 (1951)

\bibitem{Perc1} G~Baym: ``Confinement of quarks in nuclear matter,''
Physica A {\bf 96}, 131 (1979)

\bibitem{Perc2} T.~{\c C}elik, F.~Karsch and H.~Satz:   ``A percolation
approach to strongly interacting matter,''
Phys. Lett. B {\bf 97}, 128 (1980)


\bibitem{HS}
See e.g. H. Satz: {\it Extreme States of Matter in Strong Interaction
Physics: An Introduction}, Lect. Notes Phys. {\bf 841} (2012), and
references therein

\bibitem{first1}
 L.D. McLerran and B. Svetitsky:   ``A Monte Carlo Study of SU(2)
Yang-Mills Theory at Finite Temperature,''
Phys. Lett. B {\bf 98}, 195 (1981).

\bibitem{first2}
J. Kuti, J. Polonyi and K. Szlachanyi:   ``Monte Carlo Study of
SU(2) Gauge Theory at Finite Temperature,''
Phys. Lett. B {\bf 98}, 199 (1981).

\bibitem{first3}
J. Engels, F. Karsch, I. Montvay and H. Satz:  ``High Temperature
SU(2) Gluon Matter on the Lattice,''
 Phys. Lett. B {\bf 101}, 89 (1981).

\bibitem{review} F. Karsch:   ``Lattice QCD at high temperature and density,''
Lect. Notes. Phys. {\bf 583}, 209 (2002)

\bibitem{fodor} Y.~Aoki, G.~Endrodi, Z.~Fodor, S.~D.~Katz and K.~K.~Szabo:
  ``The Order of the quantum chromodynamics transition predicted by the
standard model of particle physics,''
  Nature {\bf 443}, 675 (2006)

\bibitem{fk} A.~Bazavov {et al.}  [HotQCD]:
  ``Equation of state in ( 2+1 )-flavor QCD,''
  Phys. Rev. D {\bf 90}, 094503 (2014)

\bibitem{C1}
S. Ejiri, F. Karsch and K. Redlich:
  ``Hadronic fluctuations at the QCD phase transition,''
  Phys. Lett. B {\bf 633}, {275}  (2006)

\bibitem{C2} A. Bazavov, et al., [HotQCD Collaboration]:
  ``Fluctuations and Correlations of net baryon number, electric charge,
and strangeness: A comparison of lattice QCD results with the hadron
resonance gas model,''
Phys. Rev.  D {\bf 86}, {034509}   (2012)

\bibitem{C3}  A.~Bazavov, H.-T.~Ding, P.~Hegde, O.~Kaczmarek, F.~Karsch,
E.~Laermann, Y.~Maezawa and S.~Mukherjee {\it et al.}:
  ``Additional Strange Hadrons from QCD Thermodynamics and Strangeness
Freezeout in Heavy Ion Collisions,''
{\bf 113}, 072001 (2014)

\bibitem{taw1}
F.~Karsch, K.~Redlich and A.~Tawfik:
 ``Thermodynamics at nonzero baryon number density: A Comparison of lattice
and hadron resonance gas model calculations,''
Phys. Lett. B {\bf 571}, 67  (2003).

\bibitem{taw2}
F.~Karsch, K.~Redlich and A.~Tawfik:
``Hadron resonance mass spectrum and lattice QCD thermodynamics,''
  Eur. Phys. J. C {\bf 29}, 549  (2003).

\bibitem{frit}
  F.~Karsch:  ``Thermodynamics of strong interaction matter from lattice
QCD and the hadron resonance gas model,''
 Acta Phys. Polon. Supp.  {\bf 7},   117 (2014)

\bibitem{pdg}
K. A. Olive, et al., [Particle Data Group]: Chin. Phys. C {\bf  38}, 090001
(2014)

\bibitem{redb} For a  review, see eg.
P. Braun-Munzinger, K. Redlich, and J. Stachel: ``Particle
Production in Heavy Ion Collisions'',   In  Hwa, R.C. (ed.):
{\it Quark Gluon Plasma 2}, World Scientific, (2004) p491

\bibitem{tc}  T.~Bhattacharya, M.~I.~Buchoff, N.~H.~Christ, H.-T.~Ding,
R.~Gupta, C.~Jung, F.~Karsch and Z.~Lin {\it et al.}:
  ``QCD Phase Transition with Chiral Quarks and Physical Quark Masses,''
Phys. Rev.  Lett. {\bf 113}, 082001 (2014)



\bibitem{jc}
 J.~Cleymans and H.~Satz:
  ``Thermal hadron production in high-energy heavy ion collisions,''
  Z. Phys. C {\bf 57}, 135 (1993)

\bibitem{redf}  K.~Redlich, J.~Cleymans, H.~Satz and E.~Suhonen:
  ``Hadronization of quark - gluon plasma,''
Nucl. Phys. A {\bf 566},  391 (1994)

\bibitem{heppe}
  P.~Braun-Munzinger, I.~Heppe and J.~Stachel:
  ``Chemical equilibration in Pb + Pb collisions at the SPS,''
  Phys. Lett. B {\bf 344}, 43 (1995)

\bibitem{beca} F. Becattini:   ``A Thermodynamical approach to hadron
production in e+ e- collisions,''
 Z. Phys. C {\bf 69}, 485 (1996)



\bibitem{andronic1}
A. Andronic:  ``An overview of the experimental study of
quark-gluon matter in high-energy nucleus-nucleus collisions,''
 Int. J. Mod. Phys. A {\bf 29}, 1430047 (2014).

\bibitem{andronic2}
J.~Stachel, A.~Andronic, P.~Braun-Munzinger and K.~Redlich:
  ``Confronting LHC data with the statistical hadronization model,''
  J.\ Phys.\ Conf.\ Ser.\  {\bf 509}, 012019  (2014)

\bibitem{privat} J. Rafelski: private communication

\bibitem{rafeldan} J.~Rafelski and M.~Danos:
 ``The Importance Of The Reaction Volume In Hadronic Collisions,''
  Phys. Lett.  B {\bf 97}, 279 (1980).

  \bibitem{tounsi}
 S.~Hamieh, K.~Redlich and A.~Tounsi:
  ``Canonical description of strangeness enhancement from p-A to Pb Pb
collisions,''
  Phys. Lett. B {\bf 486}, 61 (2000).


\bibitem{oeschler}

  J.~Cleymans, H.~Oeschler and K.~Redlich:
  ``Influence of impact parameter on thermal description of relativistic
heavy ion collisions at (1-2) A-GeV,''
  Phys. Rev. C {\bf 59}, 1663 (1999).

  \bibitem{turko1}
   K.~Redlich and L.~Turko:
  ``Phase Transitions in the Statistical Bootstrap Model with an Internal
Symmetry,''
  Z. Phys. C {\bf 5}, 201 (1980).

  \bibitem{turko2}
   L.~Turko:
  ``Quantum Gases With Internal Symmetry,''
  Phys. Lett. B {\bf 104}, 153 (1981).

  \bibitem{petrov}
   M.~I.~Gorenstein, V.~K.~Petrov and G.~M.~Zinovev:
  ``Phase Transition in the Hadron Gas Model,''
  Phys. Lett. B {\bf 106},  327 (1981)

 \bibitem{rafm}
  B.~Muller and J.~Rafelski:
  ``Role of Internal Symmetry in $p \bar{p}$ Annihilation,''
  Phys. Lett. B {\bf 116}, 274 (1982)

\bibitem{hagedornred} R.~Hagedorn and K.~Redlich:
``Statistical Thermodynamics in Relativistic Particle and Ion Physics:
Canonical or Grand Canonical?,''
Z. Phys. C {\bf27}, 541 (1985)

\bibitem{Derreth}
  C.~Derreth, W.~Greiner, H.~T.~Elze and J.~Rafelski:
  ``Strangeness Abundances In Anti-p Nucleus Annihilations,''
  Phys.\ Rev.\ C {\bf 31}, 1360 (1985)

\bibitem{Elze}
  H.~T.~Elze, W.~Greiner and J.~Rafelski:
  ``Color Degrees of Freedom in a Quark - Glue Plasma at Finite Baryon
Density,''
  Z.\ Phys.\ C {\bf 24}, 361 (1984)


\bibitem{s1}
  F.~Becattini, P.~Castorina, J.~Manninen and H.~Satz:
  ``The Thermal Production of Strange and Non-Strange Hadrons in e+ e-
Collisions,''
  Eur. Phys. J. C {\bf 56}, 493 (2008)

\bibitem{b1}
  A.~Andronic, F.~Beutler, P.~Braun-Munzinger, K.~Redlich and J.~Stachel:
  ``Thermal description of hadron production in e+e- collisions revisited,''
  Phys. Lett. B {\bf 675}, 312 (2009)


\end{thebibliography}
\end{document}